\documentclass{emulateapj}
\usepackage{epsf}
\usepackage{bm}
\begin{document}
\title{The mass of first stars}
\author{Hajime Susa\altaffilmark{1}}
\affil{Department of Physics, Konan University, Okamoto, Kobe, Japan}
\altaffiltext{1}{susa@konan-u.ac.jp}
\begin{abstract}
We perform a three dimensional radiation hydrodynamics simulation
 to investigate the formation of first stars from initial
 collapse of a primordial gas cloud to formation and growth of
 protostars. 
 The simulation is integrated until $\sim$0.1 Myrs after the
 formation of the primary protostar by which the protostars have 
 already settled onto main sequence stars. 
This is the first attempt of simulating first star formation to take into account the
 ultraviolet radiative feedback effect by the multiple protostars as well as the
 three dimensional effects such as fragmentation of the accretion disk.
We find that the mass accretions onto the
 population III protostars are significantly suppressed by the radiative feedback
 from themselves. As a result, we find five stars formed in this
 particular simulation, and that the final mass of the stars are
 $\la 60M_\odot$, including a star of $4.4M_\odot$. 
Formation of such a star hints at the existence of even lower-mass
 stars that would live today.
\end{abstract}

\section{Introduction}
Formation of first stars has been investigated intensively 
in the last decade mainly from theoretical aspects.
Following the theoretical predictions, the first stars form in the
mini-halos of mass $\sim 10^5-10^6M_\odot$ \citep{haiman96,
tegmark97, nishi_susa99,fuller_couchman00,abel02,bromm02,yoshida03}.

The ingredient of first star formation is the primordial gas, which does
not contain heavy elements or cosmic dusts. Because of the lack of these
efficient coolants, the primordial gas cools
inefficiently especially at low temperatures ($T \la
10^4$K). Therefore, the gas is kept relatively warm ($\sim 10^3$K) while it
collapses to form stars, in contrast to the case of the interstellar gas,
whose temperature is $\sim 10$K during the present-day star formation for $n_{\rm H} \la
10^{10}{\rm cm^{-3}}$ \citep[e.g.][]{omukai00}. 
As a result, the gravitationally collapsing primordial clouds
are very massive $\sim 1000M_\odot$, since they have
to be more massive than the Jean's mass, which is proportional to $T^{3/2}$. 
In addition, formation of such a massive
prestellar core leads to huge mass accretion rate onto the
protostar in the later mass accretion phase\cite[e.g.][]{omukai98,bromm99,abel00,NU01,abel02,yoshida06}.
Following these theoretical evidences, most of the first stars were once expected to
be very massive ($\ga 100M_\odot$).

On the other hand, the studies on the mass accretion phase have advanced
recently, which revealed that the first star formation process
seems to be more complicated than expected before\citep{turk09,clark11a,clark11b,smith11,greif11,greif12}. 
They found that a heavy disk formed around the primary protostar,
because of the angular momentum of the prestellar core, gained by
the tidal interactions with other cosmological overdense regions.
The heavy disk fragments into small pieces since it is gravitationally
unstable. As a result, a ``star cluster'' could be formed instead of a
single very massive star. 

These results seem to be robust until the
primary protostar grows to $\ga 20 M_\odot$. 
After the mass of the protostar exceeds $\sim 20 M_\odot$, significant
ultraviolet radiation flux will be emitted from the protostar\citep{omukai03,hosokawa09,hosokawa11}. 
Thus, the ultraviolet radiation from the protostar significantly affects 
the later evolution of the system.

\citet{hosokawa11} have addressed this feedback effect directly by two
dimensional radiation hydrodynamics simulations. They found that
the accretion disk around the primary protostar is photoevaporated due to
the radiative feedback, followed by the rapid decline of the mass
accretion rate onto the protostar. The final mass of the protostar in
their simulation is 43 $M_\odot$.

\citet{stacy12} also tried to assess the final mass of the first stars
in their three dimensional cosmological calculations including the fragmentation of
the disk as well. They also found that the ultraviolet radiative feedback strongly suppress
the mass accretion onto the protostars. However, the integrated physical
time is $\sim 5000$yrs, which is too short to predict the final mass of
the first stars, since it will take $\sim 0.1$Myrs until the protostar
settles onto the main sequence\citep[e.g.][]{hosokawa09}.

In this paper, we report the results of three dimensional radiation
hydrodynamics simulations on the formation of first stars that follow the
evolution of the system for 0.1 Myrs after the formation of the primary protostar.
We take into consideration the three dimensional effects, as well as
the radiative feedback from the protostars.

\section{Numerical simulation}
\label{simulation}
We study the formation of first stars by radiation hydrodynamics
simulations. 
 We employ the Bonner-Ebert sphere of $n_{\rm
H}=10^4{\rm cm^{-3}}, T=200{\rm K}$ as the initial condition of the
simulation. This initial condition is motivated by the cosmological
simulations\citep{abel02,yoshida03} in which such clouds are found in
minihalos of mass $10^5-10^6M_\odot$, at the ``loitering'' phase. The
loitering phase corresponds to the epoch when the cloud becomes
quasi-static, because H$_2$ line cooling for $n_{\rm H} \ga 10^4{\rm
cm^{-3}}$ becomes less efficient than that for $n_{\rm H} \la 10^4{\rm
cm^{-3}}$.
In order to make the gas cloud slightly gravitationally unstable, we
increase the density by 20\%. As a result, the total mass of the cloud
is $2600M_\odot$. We also add uniform rotation
to the gas sphere with $\Omega=2\times 10^{-14}{\rm rad/s}$. The angular
velocity of this value results in very similar specific angular momentum
distribution to that found in the cosmological simulations by
\citet{yoshida06} at the final stage
of the run-way collapse phase.

\begin{deluxetable}{clc}
\tabletypesize{\scriptsize}
\tablecaption{Chemical reactions}
\tablewidth{0pt}
\tablehead{
\colhead{number} & \colhead{reaction} & \colhead{reference}
}
\startdata
1 & H$^+$ + e$^-$ $\rightarrow$ H + $\gamma$  & SP \\
2 & H + e$^-$ $\rightarrow$ H$^-$ + $\gamma$  & GP \\
3 & H + H$^-$ $\rightarrow$ H$_2$ + e$^-$  & GP \\
4 & 3H $\rightarrow$ H$_2$ + H$^-$  & PSS \\
5 & H + H$_2$ $\rightarrow$ 3H  & SK \\
6 & 2H + H$_2$ $\rightarrow$ 2H$_2$ & PSS \\
7 & 2H$_2$ $\rightarrow$ 2H+H$_2$   & SK \\
8 & H + e$^-$ $\rightarrow$ H$^+$ + 2e$^-$  & SK \\
9 & 2H $\rightarrow$ H+ H$^+$ + e$^-$  & PSS \\
10 & H + H$^+$ $\rightarrow$ H$_2^+$ + $\gamma$  & GP \\
11 & H$_2^+$ + H $\rightarrow$ H$_2$ + H$^+$  & GP \\
12 & H$_2$ + H$^+$ $\rightarrow$ H$_2^+$ + H  & GP \\
13 & H$^+$ + H$^-$ $\rightarrow$ H$_2^+$ + e$^-$  & GP \\
14 & H$_2^+$ + e$^-$ $\rightarrow$ 2H   & GP \\
15 & H$^-$ + e$^-$ $\rightarrow$ H + 2e$^-$   & SK \\
16 & H$^-$ + H$^+$ $\rightarrow$ 2H   & GP \\
17 & H$_2$ + e$^-$ $\rightarrow$ H + H$^-$   & GP \\
18 & H$_2$ + e$^-$ $\rightarrow$ 2H +e$^-$   & GP \\
19 & H + $\gamma$ $\rightarrow$ H$^+$ + e$^-$   & SU \\
20 & H$_2^*$ + $\gamma$ $\rightarrow$ 2H   & KBS+WHB \\
21 & H$_2^+$ + $\gamma$ $\rightarrow$ H + H$^+$   & TG  \\
22 & H$^-$ + $\gamma$ $\rightarrow$ H + e$^-$  & ST
\enddata


\tablecomments{References.SP:\citet{spitzer78}, GP:\citet{GP98},
 SK:\citet{SK87}, PSS:\citet{PSS83}, SU:\citet{susa06},
 KBS:\citet{kepner97},ST:\citet{stancil94}, WHB:\citet{wolcott-green11},TG:\citet{tegmark97}}
\label{reactions}

\end{deluxetable}

We use the code Radiation-SPH \citep{susa04,susa06} in order to solve
the equations of hydrodynamics, non-equilibrium primordial chemistry of
six species, e$^-$, H$^+$, H, H$_2$, H$^-$, H$^+_2$, and radiation
transfer of ultraviolet photons. The reactions included in the code
are listed in Table \ref{reactions}. Transfer of ultraviolet radiation
is assessed by ray-tracing scheme, in which  neighbor SPH particles are
connected to make up the rays \citep{susa06}.
Then we calculate the optical
depth at the Lyman limit as well as the H$_2$ column density by the
ray-tracing scheme. Using the optical depth at the Lyman limit, we can
assess the correct photoionization/photoheating rate by integrating the
spectrum before we start the simulation\citep[e.g.][]{susa04}, as far as 
we employ on-the-spot approximation, that is assumed in this simulation. H$_2$
column density is used to calculate the self-shielding function of
Lyman-Werner (LW) photons. 
We use updated self-shielding function for LW radiation transfer\citep{wolcott-green11}.
H$_2^+$ photodissociation is
also taken into account, based upon the cross-section in
\cite{stancil94}. H$^-$ radiative detachment is assessed using the
fitting formula for the cross-section in \citet{tegmark97}.
We also assume the radiation below LW band is optically thin.

We include the cooling processes of primordial gas such as H/H$_2$ line
cooling, H$_2$ formation heating/dissociation cooling, H
ionization/recombination cooling, bremsstrahlung, optically thin H$^-$
cooling and collision induced
emission (CIE) cooling\footnote{CIE is added just for completeness,
since it is important only above $10^{14}{\rm cm^{-3}}$
\citep{yoshida08}.}. 
We take into account the shielding of H$_2$ line
emission, utilizing the shielding function proposed by \citet{ripamonti}.  
In the present version of RSPH code, we update the hydrodynamics
gravity, and radiative transfer at Courant time, while the energy
equation and chemical reaction equations are integrated at smaller time
step.

The mass of an SPH particle in the present work is set to be $m_{\rm SPH}=4.96\times
10^{-3}M_\odot$ and the number of neighbor particles is $N_{\rm neib}=50$, that correspond to the mass resolution of $M_{\rm
res}=2N_{\rm neib}m_{\rm SPH} = 0.496 M_\odot$\citep{bate_burkert97}. This
mass resolution is equivalent to the Jeans mass of $n_{\rm
H}=10^{12}{\rm cm^{-3}}, T=300{\rm K}$, and comparable to that in
\citet{stacy12}. 

In order to trace the evolution in mass accretion phase, we employ sink
particles. If the density at an SPH particle exceeds $n_{\rm
sink}=3\times10^{13}{\rm cm^{-3}}$, we change the SPH particle into a sink
particle. In addition, if SPH particles fall
within the sphere with radius of  $r_{\rm acc}=30{\rm AU}$ centered on a sink
particle, and they are gravitationally bound with each other, these
SPH particles are merged to the sink particle, conserving the linear
momentum and mass. The accretion radius $r_{\rm acc}$ is again
comparable to that employed in \citet{stacy12}. 
We remark that the sink-sink merging is not allowed in the present
numerical experiment, since the radius of the protostar is less than $\sim
1$AU\citep{hosokawa09}, which is much smaller than the employed accretion radius $r_{\rm acc}$. 
We regard the mass of the sink particles as the mass of protostars. 
We also assume that the sink particles do not push
surrounding SPH particles, i.e. the pressure forces from sink particles
to surrounding SPH particles are omitted. The recipe of sink particles
employed in the present work is known to overestimate the mass accretion
rate\citep[e.g.][]{bate95,bromm02,martel06}. 
In addition, the employed
accretion radius $r_{\rm acc}$ is 30AU, which is much larger than the
radius of protostars\citep{hosokawa09}. Thus, we have to keep in mind that
resultant mass of the formed sink particles would be larger than the actual
mass of first stars.
We also remark that we cut the central spherical region with radius
0.6pc out of the cloud, just after the formation of the first sink in order to save the computational time.  The outer envelope of $r > 0.6$pc hardly affect the inner region within $10^5$yrs.

We turn on the sinks(stars) when they are created, 
although the initial mass of them are so small that very little UV
photons are emitted initially.
The mass accretion rates onto sinks in the present simulation are obtained by
averaging over $10^3$yrs in order to avoid artificial jaggy behaviours due to
SPH discreteness. We feed this mass accretion rate to the protostellar
evolution model to assess the stellar radius/effective
temperature.\footnote{We have to keep in mind the limitation of present
treatment of protostellar model in which steady accretion is
assumed\citep[e.g.][]{PSS86}. More violent/clumpy mass accretion could remain the effective temperature
of protostar colder\citep{smith12} to larger masses.}
The evolving luminosity and effective temperature of a protostar is
obtained based on the calculation by \citet{hosokawa09}.
They have calculated the evolution of protostars with given (fixed)
mass accretion rates. 
On the other hand, we obtain the protostellar masses ($M$) and the
mass accretion rates ($\dot{M}$) self consistently from the
hydrodynamics simulation. Then we can assess the luminosities and
effective temperatures by interpolating the data by \citet{hosokawa09} at
every time step. These luminosities and temperatures are used to give the
luminosities and black body spectra of the protostars.
Hence, the protostellar evolution model is self consistently fed to
radiation hydrodynamics calculations.

\section{Results}
\label{results}
We perform radiative hydrodynamics simulations of the first star formation
with radiative feedback. We also perform a run with no feedback for comparison.
\subsection{Fragmentation of the disk around the primary protostar}
We start the simulation from the rigidly rotating Bonner-Ebert sphere
around the loitering phase. The cloud starts to collapse in run-away
fashion, i.e. the central density keeps growing while the outer part of
the cloud is left in the envelope. As a result, density profile of
$\propto r^{-2.2}$ is built up during the run-away phase \citep{yoshida06}.

Eventually, the central density exceeds $n_{\rm sink}$, and a sink
particle is formed at the center of the cloud. The surrounding gas
starts to accrete onto the sink particle subsequently. Since the gas has
a significant amount of specific angular momentum, the accreting gas forms
an accretion disk around the sink particle. The amount of specific
angular momentum in the run-away phase is close to that of the similarity
solution, which is approximately 0.5 times the value of Kepler rotation
at the Jeans radius, i.e. the core radius \citep{yoshida06}. Thus, the
radius of the disk is 0.25 times smaller than their original radius at the run-away phase, since the
centrifugal force is proportional to the square of the specific angular momentum.

\begin{figure*}
\plotone{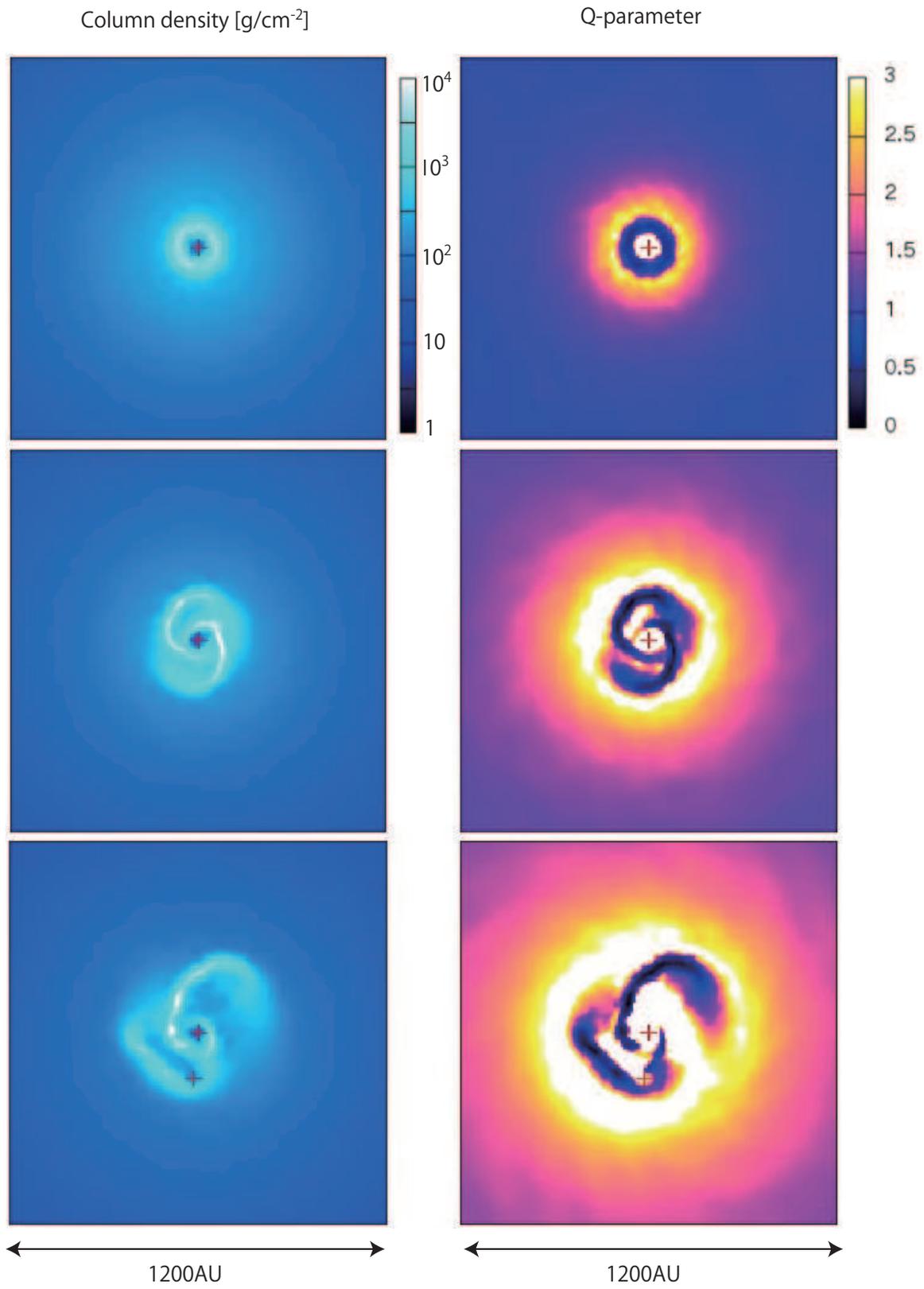}
\caption{{Face-on} view at the beginning of the accretion phase. Left
 column: Three snapshots (320yr, 620yr, 860yr from top to bottom) of
 color contour of the gas column density. Right column: Toomre's Q-parameter
 at the same moment.}
\label{fragmentation}
\end{figure*}

After the formation of the small gas disk around the first sink, the gas
keeps accreting onto the disk. As a result, the mass and the radius of
the disk increase. At the same time, the temperature of the disk
decreases by the radiative cooling. The left column of
Fig.\ref{fragmentation} shows the face-on views of the disk column
density at three epochs corresponding to 320yrs, 620yrs and 860yrs after the formation of
the first sink. The red crosses denote the position of sink particles.

In the early epoch of the accretion phase, a smooth disk forms around
the sink particle (top panel), followed by the formation of spiral
arms (middle panel) and the fragmentation of the arms (bottom panel). We
can also find a few high column density peaks in the bottom panel. In
fact, next sink particles are born from these peaks within a few
hundred years. 

The right column of Fig.\ref{fragmentation} shows the color contour of
Toomre's $Q$-parameter, which is given as
$$Q\equiv c_{\rm s}\Omega_{\rm orb}/(\pi G \sigma)$$
in case we assume Keplerian motion. Here $c_{\rm s}$ denotes the sound
velocity of gas, $\Omega_{\rm orb}$ is the angular velocity around the
central sink particle and $\sigma$ is the column density of the disk.
In the case that the $Q$-parameter is less than unity, a smooth disk with density
perturbations becomes gravitationally unstable. In fact, the $Q$-parameter of
the disk at the early phase (top) is already less than unity, so the disk is
unstable (middle and bottom).
 
The time scale of disk instability is given by the linear perturbation
theory \citep{toomre64},
which is given as $\Omega_{\rm orb}^{-1}(Q^{-2}-1)^{-1/2}$. 
The typical
value of $Q$-parameter in the disk at early phase (top) is $\sim 0.5$,
and the angular velocity is $\Omega_{\rm orb}\simeq 10^{-9}{\rm s^{-1}}$.
Thus  the perturbation growth time scale is $\sim 20$yrs, which is
comparable to the time scale of the generation of the spiral structure.
Thus, spiral structures seems to develop due to the gravitational instability,
which could be understood as Toomre's criteria.
On the other hand, it takes several hundred years for another sink to be
born (bottom), and seemingly via the fragmentation of
the spiral arms. Therefore, formation of the sink particles in the disk cannot be
understood solely by the simple $Q$-parameter argument above, but requires non-linear
calculations.

\begin{figure*}
\plotone{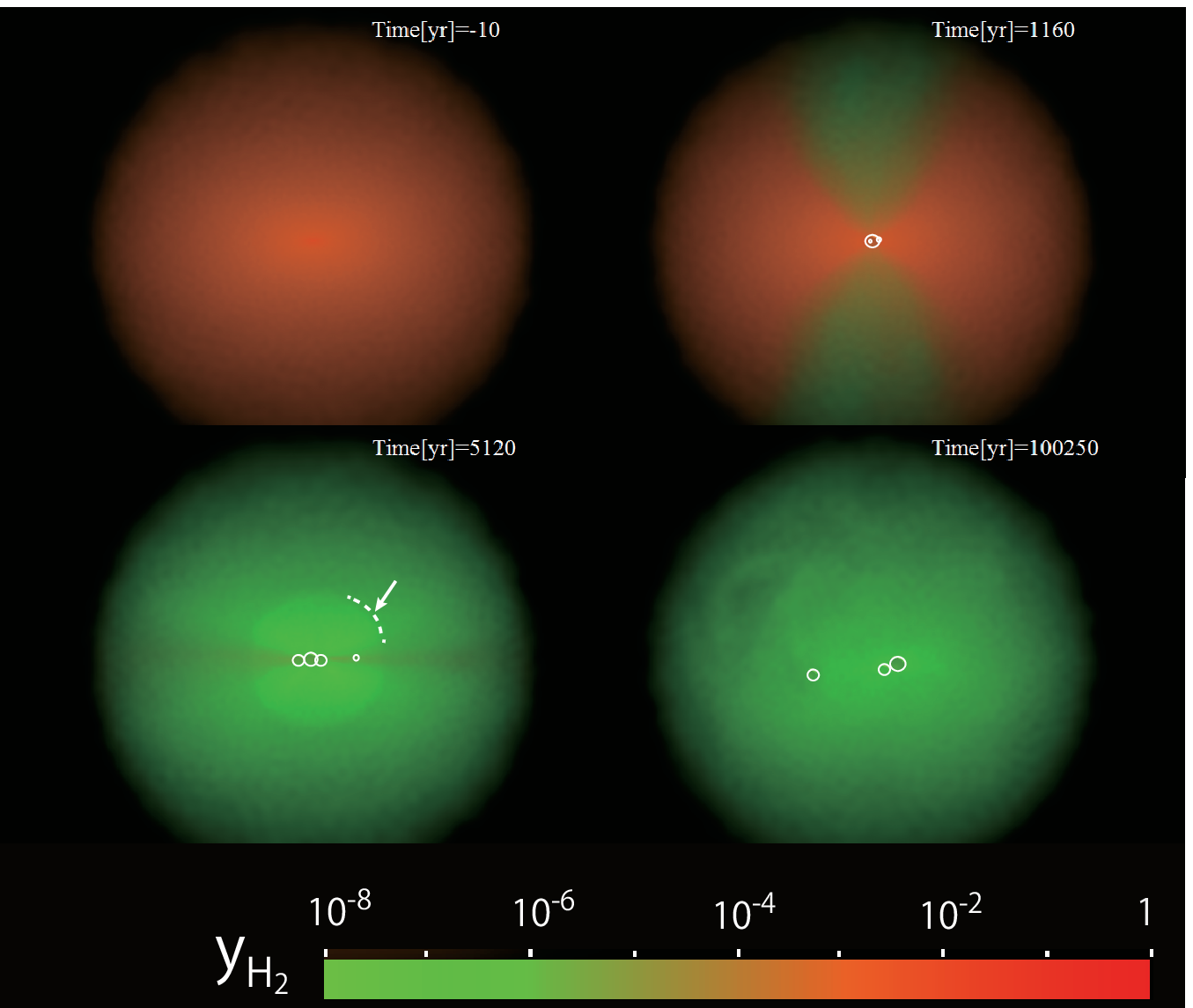}
\caption{Edge-on views of gas distribution inside $r<10^4$AU (0.05pc) at
 four snapshots. {Top row:
 from left to right, $t=$-10yr,1160yr. Bottom row:$t=$ 5120yr,
 100250yr. $t$ represents the time after the first sink
 formation. Color shows the H$_2$ fraction, and the small spheres with
 white rim}
 represent the positions of sink particles. White arrow and dashed curve
 in the bottom left panel denotes the position of the shock front.}
\label{h2map}
\end{figure*}

\subsection{Effects of radiative feedback}

Fig.\ref{h2map} illustrates the evolution of the edge-on view of the
central $10^4$AU (0.05pc) in radius. The color shows the fraction of H$_2$ molecules,
and the transparency denotes the gas density. Small spheres are the
position of sink particles. Initially, H$_2$ fraction is quite high
($y_{{\rm H}_2}\sim 10^{-2}$, upper left panel). Eventually, the polar
region is photodissociated as the sink particles grow (top right). Some H$_2$ rich
regions remain along the equatorial plane due to the self-shielding
(bottom left), but
finally they disappear after 0.1Myrs (bottom right).

\begin{figure}
\plotone{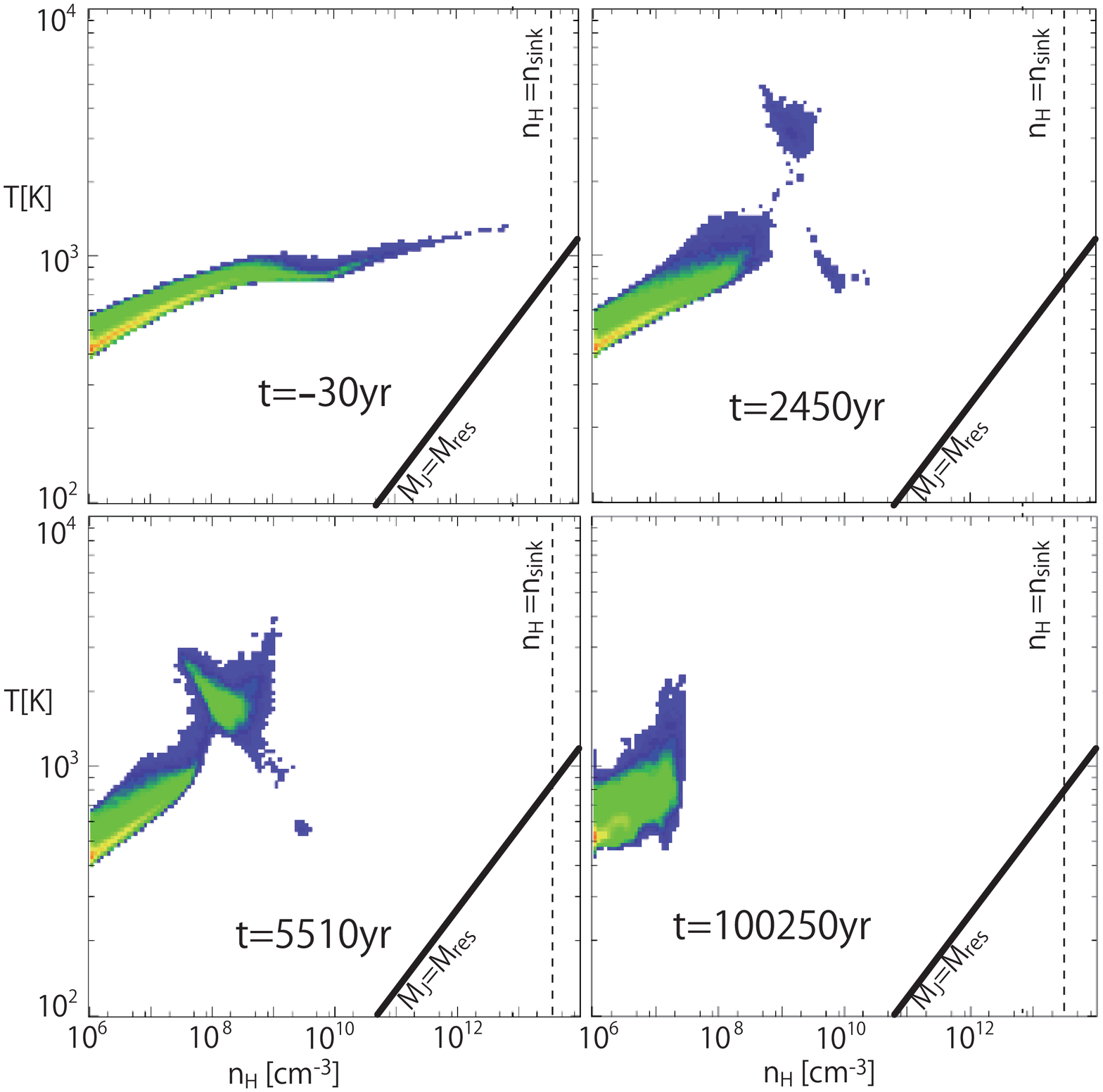}
\caption{Four snapshots on the density-temperature plane. Color represents the
 number of SPH particles drop on the logarithmic bin on the plane. Thick
 solid lines show the resolution limit of this simulation, while dashed
 lines represent the number density above which the SPH particles are
 converted to sink particles.}
\label{phase}
\end{figure}

\begin{figure}
\epsscale{1.3}
\plotone{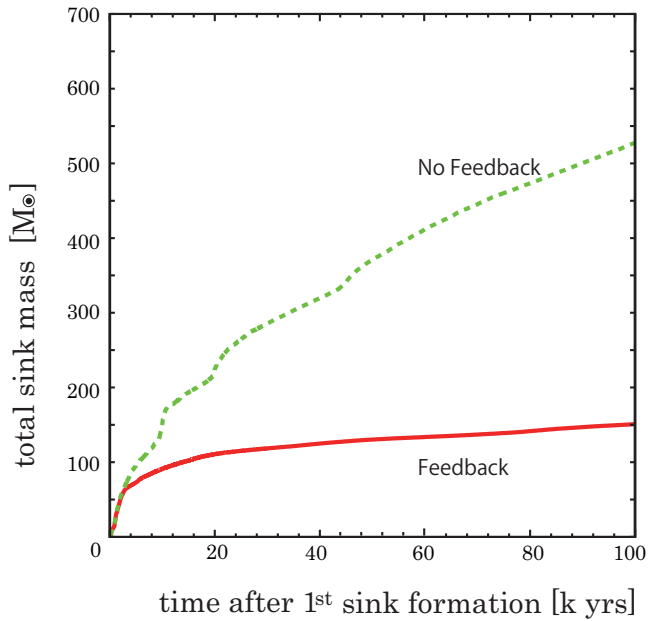}
\caption{Evolution of the total mass incorporated in sink
 particles. The solid line corresponds to the case with feedback, while
 the dashed line represents the case without feedback.}
\label{star_all}
\end{figure}

\begin{figure}
\epsscale{1.1}
\plotone{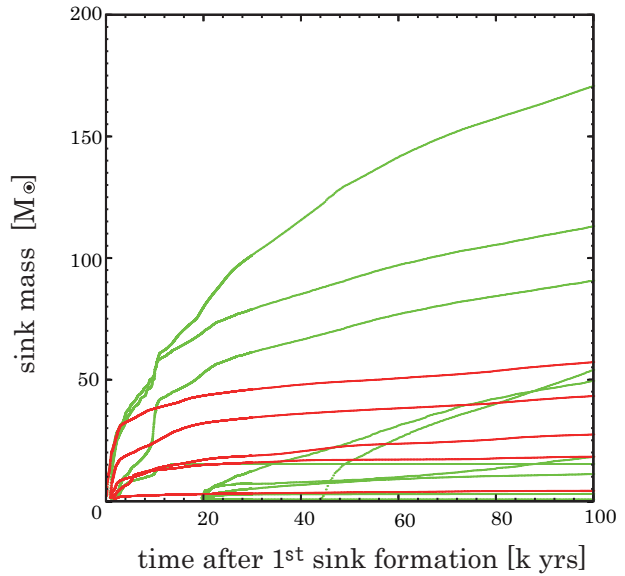}
\caption{The mass of individual sink particle is plotted as a function
 of time after the first sink formation. Red curves for the runs with
 feedback, while the green curves for the case without feedback.}
\label{star_ind}
\end{figure}

\begin{figure}
\epsscale{1.1}
\plotone{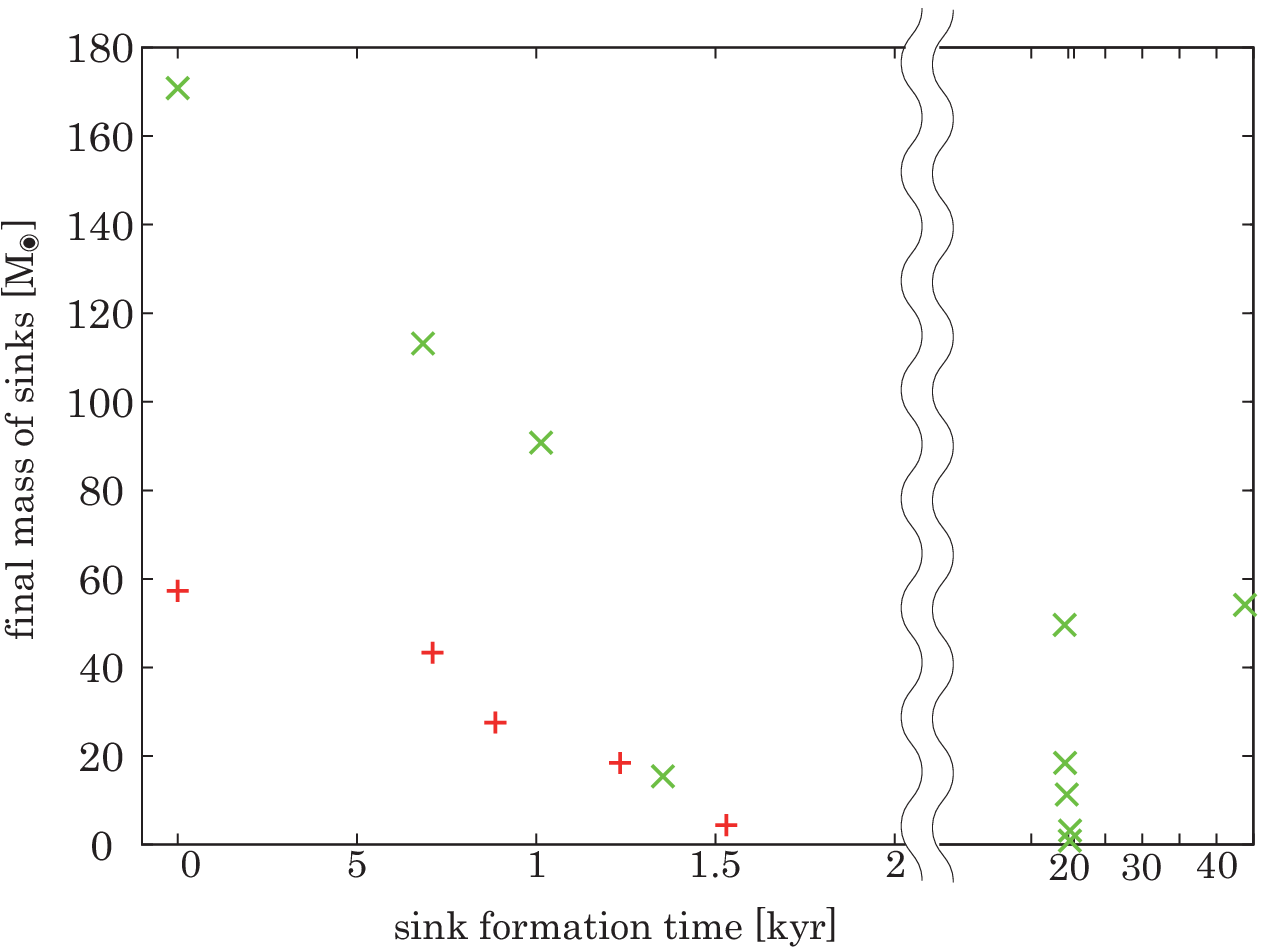}
\caption{Formation time of individual sink particle v.s. final
 mass. Red crosses : feedback run, Green vertices: no feedback run. }
\label{ftime_fmass}
\end{figure}

Fig.\ref{phase} illustrates the evolution of the system on
density-temperature plane. Color map shows the frequency distribution of SPH
particles on the plane. Four panels show the snapshots at
$t=$-10yr, 2450yr, 5510yr, 100250yr, respectively. The distribution of
SPH particles on the top left panel is similar to the well known curve
of the evolution of collapsing primordial gas in run-away phase \citep[e.g.][]{PSS83}, since it
corresponds to the epoch just before the first sink formation. After the
first sink formation, dense gas ($\ga 10^{10}{\rm
cm}^{-3}$) is split to high temperature gas ($\la 7000$K) and low
temperature gas ($\la 1000$K, top right panel). The former corresponds to the radiatively heated gas and the shock heated gas, while the latter is the
self-shielded cold gas orbiting around the sink particles. Then, the
high temperature gas in the high density region expands due to the
increased thermal pressure, generating a shock propagating to low density region (bottom left
panel). In fact, we also have seen the shock wave in the bottom left
panel of Fig.\ref{h2map}, marked by white dashed curve. 
Finally, the dense cold gas disappears (bottom right), which means that star formation no longer proceeds.

The solid line in Fig.\ref{star_all} shows the time evolution of the total
mass in the sink particles in the feedback run, whereas the dashed line represents that
without radiative feedback effects. 
It is clear that the mass accretion onto the sink particles is highly suppressed by the radiative
feedback. The total sink mass of the run with feedback at the end of the simulation ($\sim$0.1 Myr) is
less than  a third of that without feedback.
{We also find that the mass accretion rate in feedback run is
smoother than that of no feedback run. This is because the gas in the
latter case is more clumpy than that of the former, due to the absence
of additional heating processes provided by the ultraviolet radiation
from the protostar.}
Fig.\ref{star_ind} illustrates the time evolution of the mass of each sink particle.
The red curves represent those in the run with feedback, while the green
curves are those results without feedback. In the feedback run, we have one 
star more massive than 50$M_\odot$ at $0.1$Myr ($\sim 57 M_\odot$),
whereas we have three stars in the range of $50-200 M_\odot$ in the no feedback run. 

The minimal sink mass in the feedback run is
4.4$M_\odot$, and the total number of the sink particles is 5. On the
other hand, in the no feedback run, the minimal sink mass is
0.84$M_\odot$, and the total number of sink particles is 10. It is also worth noting that fragmentation of the disk in the no
feedback case continues until much later time ($\sim 4\times 10^4$yr)
than that in the feedback run ($\la 1500$yr), because the molecular rich
disk is not destroyed until much later phase in the no feedback run.
The difference in the number of sink particles and the minimal mass might
come from such effect. However, it is premature to draw definitive
conclusion on this issue, since our results are based upon only a single
realization.

{Fig.\ref{ftime_fmass} plots the sink formation time and the final
mass (at $10^5$ yr after 1st sink) for all sink particles. Red crosses
denote the sinks found in feedback run, while the green vertices are in
the no feedback run. Basically, earlier formation leads to more massive
sinks because massive sinks gather gas more efficiently than less massive
ones. This trend is clear in the feedback run, while some different
behaviours are found in no feedback case. In the latter case, the gas
distribution is more clumpy, which allows the sink formation at late
epochs such as 20kyr or 40kyr after the first sink formation, as
mentioned in the previous paragraph. In such
cases,  the growth of sinks is not affected by the sinks formed at the first
episode, because it proceeds at relatively spatially distant position from the first sinks.}

\subsection{Ejections}
\begin{center}
\begin{figure}
\epsscale{1.0}
\plotone{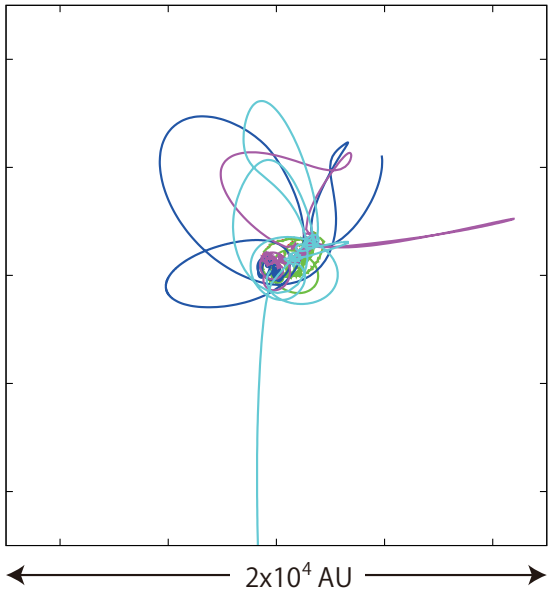}
\caption{The trajectories of sink particles in edge-on view.}
\label{trajectory}
\end{figure}
\end{center}

In the simulation with feedback, we find a sink particle 
is kicked away
from the central dense region via the gravitational N-body interaction,
so-called ``slingshot'' mechanism. 
Fig.\ref{trajectory} shows the trajectories of all the 
sink particles within $(2\times 10^4{\rm AU})^3$ box at the central region. It is
clear that one escaping sink goes away from the central region (to the
bottom of the panel), whereas the others
remain within the box. Such a phenomenon has already been reported by
{other }groups without radiative feedback effects\citep[e.g.][]{smith11}. Thus, we confirm the
theoretical existence of such escapers also in our numerical model with
radiative feedback. The velocity of this escaping sink is $\sim 4 {\rm
km/s}$ at 0.1 pc distant from the center of the cloud, which is marginal
to escape from the host minihalo of $10^6M_\odot$.

The mass accretion onto this escaping particle almost stops after the
ejection. Consequently, the mass of the sink is $4.4M_\odot$,
that are much smaller than the ``conventional'' first stars of mass $\ga
100M_\odot$. It is also worth noting that the orbit of the other sinks are
excited by the N-body interaction with each other, although they are not
ejected. Thus, some of the sinks going through relatively low density
regions, which results in low mass accretion rate onto these sinks. 

We also remark that a star less massive than $0.8M_\odot$
is found in our higher resolution run ($M_{\rm res}=0.1M_\odot$)
with feedback, although the integrated physical time is $\sim 2\times
10^4$yrs \citep{umemura12}.  
Considering the fact that the mass resolution and the accretion radius of the
present simulations are $\sim 0.5M_\odot$ and $30{\rm AU}$, and other
higher resolution studies with/without feedback effects report the ejection of
even lower mass stars\citep{umemura12,clark11a,clark11b,smith11,greif11,greif12}, 
we presume that low mass stars less massive than 0.8$M_\odot$ are born
among the first stars, and survive through the entire history of the universe.

\section{Discussions \& conclusion}
In the presence of radiative feedback, the gas in the neighborhood of protostars is heated up significantly.
This heating process occurs mainly through photodissociation of H$_2$
molecules: Formation of H$_2$ molecules works as a heating process of
the gas, since formation process such as 3H$\rightarrow$H$_2$+H or
H$^-$+H $\rightarrow$ H$_2$ + e$^-$ releases the latent heat. In the
absence of radiative dissociation process, collisional dissociation
processes absorb thermal energy, which balance with the formation
heating. Thus, the net increase of H$_2$ molecules
causes effective heating of the gas, while the net decrease of H$_2$
results in cooling.  On the other hand, in the presence of
strong photodissociative radiation, it overwhelms other collisional
dissociation processes. The photodissociation process do not absorb
thermal energy, since the energy required to dissociate H$_2$ molecules
 is supplied by the radiation. Therefore, H$_2$
dissociation is no longer a cooling process. As a result, H$_2$
formation heating proceeds without hindrance in the absence of
the counter process. {In the present simulation, strong LW radiation
field from the protostar is the source of this heating process in dense
regions. Consequently, this heating process drive the shock wave found in
Fig.\ref{h2map}, as well as the termination of mass accretion onto the protostars.}

\begin{figure*}
\epsscale{0.8}
\plotone{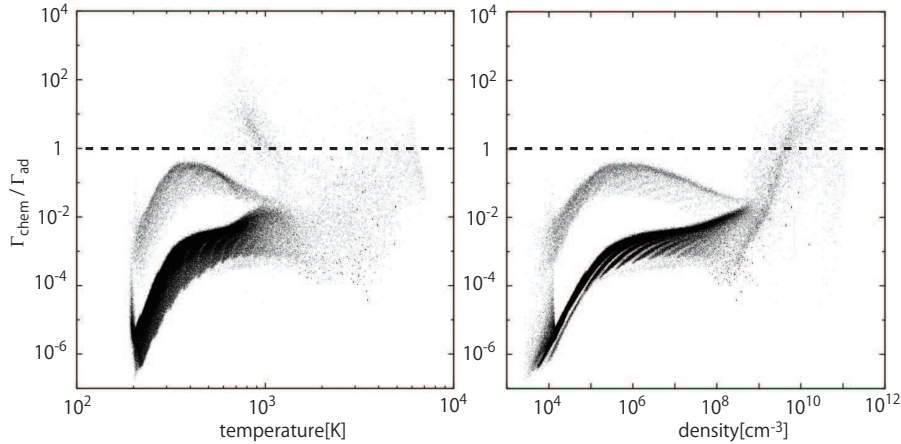}
\caption{The ratio between the chemical heating rate and the adiabatic
 heating rate is plotted as functions of temperature (left panel) and density
 (right panel) at $t=2450$yr.}
\label{heating}
\end{figure*}

In fact, Fig.\ref{heating} illustrates the ratio between
the H$_2$ formation heating rate and the adiabatic heating rate as
functions of gas temperature and density 
at $t=2450$yr. It is clear that H$_2$ formation heating is the dominant
heating process at high densities ($n_{\rm H}\ga 10^{10}{\rm
cm^{-3}}$, right panel). The temperature of these chemo-heated high
density regions is mainly around 1000K,  and the chemical heating is also
important for high temperature region at $3000{\rm K} \la T\la 7000{\rm K}$ (left panel).
It is also clear that chemical heating rate is not negligible compared to
the adiabatic heating rate even at lower densities ($\sim 10^6{\rm
cm^{-3}}$) and lower temperatures ($T\sim 300$K), which corresponds to
the dissociated polar regions. Thus, chemical heating of H$_2$ formation
play important roles on the dynamical evolution of this system.

\begin{figure}
\plotone{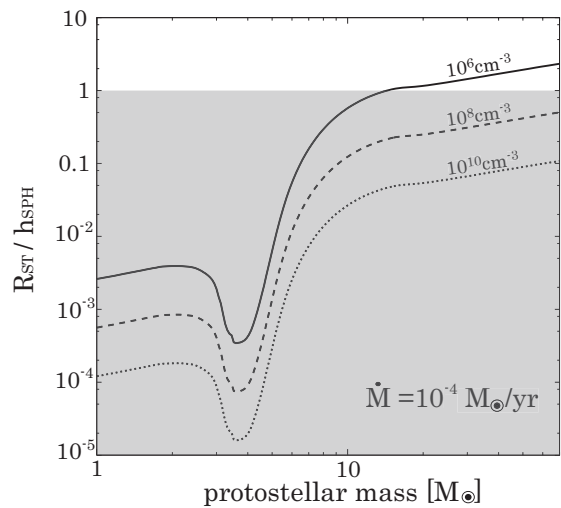}
\caption{The ratio between the Str\"{o}mgren radius $R_{\rm ST}$ and the SPH spatial
 resolution $h$ at given densities. Three curves correspond to 
$n_{\rm H}=10^{10}{\rm cm}^-3,10^{8}{\rm cm}^-3,10^{6}{\rm
 cm}^-3$. $R_{\rm ST}$ is not resolved in the shaded region.}
\label{rst_h}
\end{figure}

Heating through photoionization is also important. In fact, similar
calculation in 2D by \citet{hosokawa11}, photoionization is the dominant
heating process. They found {the breakout of ionization fronts into the
polar region}, and the gas is highly
ionized and heated up to $> 10^4$K. Finally, the disk is
photoevaporated mainly due to the photoheating process through
ionization. In the present calculation, however, ionization is not the
dominant heating process. The reason simply comes from the fact that the
spatial resolution of the present simulation is not enough to resolve
the ``initial'' Str\"{o}mgren sphere\citep{spitzer78}. 
{In Fig.\ref{rst_h}, we plot the ratio between the Str\"{o}mgren radius and the SPH spatial
 resolution at given densities. Here the mass accretion rate is
 assumed to be $10^{-4}M_\odot/yr$, which is a typical value in present
 calculation at 10kyrs-100kyrs after the first sink formation. 
It is clear that if
 the gas densities in the neighbor of the protostar is larger than $\sim
 10^6{\rm
 cm}^{-3} $, the initial Str\"{o}mgren radius cannot be resolved by the
 resolution of the present simulation. 
In the present simulation, gas particles in the neighbor of the source
stars are always denser than $10^7{\rm cm^{-3}}$, which is too high to
capture the propagation of ionization fronts.
Hence} the SPH particles
{only} in the very vicinity of the source stars are heated by
ionization ``mildly'', $T\la 10^4$K, which cannot halt the mass
accretion onto the protostar and cannot cause the breakout of
D-type ionization front which is found in \citet{hosokawa11}. 
In other words, this radiative feedback {due to photoheating }effect
is { heavily} underestimated in the
present simulation.
 
As noticed in section \ref{simulation}, the algorithm of mass accretion
onto sink particles overestimates the mass accretion rate. Combined with
the fact that feedback is underestimated, the mass accretion rate
obtained in the present numerical experiment is overestimated at any
hand. Accordingly, the final mass of the sinks should be
regarded as an upper limit of the actual mass of first stars. If we
could perform simulations at higher resolution with more realistic
mass accretion conditions, the final mass could be smaller than
$\sim 50M_\odot$. 
We also should keep in mind that this result comes
from a numerical experiment of one realization. Thus, this upper limit
should be regarded as a guide, since a slightly different initial
condition could cause different result because of the chaotic nature of
the system.  
 On the other hand, at least one of the protostars cannot
be less massive than $M \sim 20M_\odot$,
since the radiative feedback becomes prominent only for $M\ga
20M_\odot$. Thus, the mass of the primary star among the first stars
formed in a single mini-halo will fall in the range of { 
$20M_\odot$-several$\times 10M_\odot$.}

{We also remark that the evidences of $15-50M_\odot$ population III stars have
known for several years\citep{bc05,fet05,Cayrel2004,Iwamoto2005,Lai2008,jet09b,caffau12}, 
which are consistent with the present results.
In addition, however, present results do not exclude the presence of
very massive stars of $>130 M_\odot$ which result in pair instability
supernovae, since present simulation is a result of one
realization. Search for the evidences of strong ``odd-even'' effects
known as the mark of pair instability supernovae, in
dumped Lyman-$\alpha$ systems\citep{cooke11} or metal poor stars\citep{ren12} could give the
constraint on this type theoretical experiments.
}

The initial condition of the present experiment is a rigidly rotating
Bonner-Ebert sphere. Although its angular momentum distribution just
before the sink formation is close to that of cosmological simulations,
it is not fully cosmological. In cosmological simulations, the direction
of angular momentum of the disk around the primary protostar changes
depending on the  stages, since the gas motion is more turbulent.
In addition, we need more numerical experiments starting from various
initial conditions, in order to obtain statistical quantities such as
the initial mass function (IMF) of first stars.
Thus it is important to perform fully cosmological simulations of
this sort, which is left for future works.

\hspace{1cm}

In this paper, we investigated the suppression of mass accretion onto
population III proto-stars by the radiative feedback from themselves. We performed
numerical experiment of the formation of first stars using three
dimensional radiative hydrodynamics code RSPH combined with sink
particle technique. Consequently, we find
that the mass accretion is suppressed significantly and the final outcome
is a multiple stellar system consisting of five stars of
$1-60M_\odot$. The fact that low mass stars are found in this work infer
the possible existence of first stars in the local universe, although the mass of the formed stars in our simulation is still larger than 0.8 $M_\odot$.

\bigskip
 {We thank anonymous referee for his/her careful reading and constructive comments.}
 We also thank T. Hosokawa for providing the data of protostars and K. Omukai for careful reading of the manuscript. 
This work was supported by Ministry of
Education, Science, Sports and Culture, Grant-in-Aid for Scientific Research (C), 22540295. 


\end{document}